\documentclass{acm_proc_article-sp}
\usepackage{graphicx}
\usepackage{amssymb,amsmath}

\begin{document}

\title{Throughput metrics and packet delay in TCP/IP networks}

\subtitle{[Work in progress]}

\numberofauthors{4} 
%
\author{
\alignauthor
Andrei M. Sukhov\titlenote{corresponding author}\\
       \affaddr{Samara State Aerospace University}\\
       \affaddr{Moskovskoe sh., 34}\\
       \affaddr{Samara, 443086, Russia}\\
       \email{amskh@yandex.ru}
\alignauthor 
Timur Sultanov\\
       \affaddr{Samara State Aerospace University, Togliatti branch}\\
       \affaddr{Voskresenskaya st., 1}\\
       \affaddr{Togliatti, 445000, Russia}\\
       \email{tursul@rambler.ru}
\alignauthor 
Mikhail V. Strizhov\\
         \affaddr{Samara State Aerospace University}\\
       \affaddr{Moskovskoe sh., 34}\\
       \affaddr{Samara, 443086, Russia}\\
       \email{strizhov@ip4tv.ru}
\and
\alignauthor      
Alexey P. Platonov\\
       \affaddr{Russian Institute for Public Networks} \\
       \affaddr{Kurchatova sq. 1}\\
       \affaddr{Moscow, 123182, Russia}\\
       \email{plat@ripn.net}
}

\date{20 April 2009}

\maketitle

\begin{abstract}
In the paper the method for estimation of throughput metrics like available bandwidth and end-t-end capacity is supposed. This method is based on measurement of network delay $D_i$ for packets of different sizes $W_i$. The simple expression for available bandwidth $B_{av} =(W_2-W_1)/(D_2-D_1)$ is substantiated. The number of experiments on matching of the results received new and traditional methods is spent. The received results testify to possibility of application of new model.
\end{abstract}



\section{Introduction}

Measurement of throughput metrics like available bandwidth and capacity gives a great chance to predict the end-to-end performance of applications, for dynamic path selection and traffic engineering, and select among numbers of differentiated classes of service. The throughput metric is an important metric for several applications, such as grid, video and voice streaming, overlay routing, p2p file transfers, server selection, and interdomain path monitoring.

Various real-time applications in Internet, first of all, transmission audio and video information become more and more popular, however for their qualitative transmission high-speed networks are required. The major factors defining quality of the service are: quality of the equipment (the codec and a video server) and an available bandwidth of the channel. Providers and their customers should provide a demanded available bandwidth for voice and video applications to guarantee presence of demanded services in a global network.

In this paper we define a network path as the sequence of links that forward packets from the path sender to the receiver. There are various definitions for the throughput metrics, but we will adhere to the approaches accepted in a series of papers by Dovrolis {\em et al\/}~\cite{drm,jd,pdm}.

Two bandwidth metrics that are commonly associated with a path are the capacity $C$ and the available bandwidth $B_{av}$. The capacity $C$ is the maximum IP-layer throughput that the path can provide to a flow, when there is no competing traffic load (cross traffic). The available bandwidth $B_{av}$, on the other hand, is the maximum IP-layer throughput that the path can provide to a flow, given the path's current cross traffic load. The link with the minimum transmission rate determines the capacity of the path, while the link with the minimum unused capacity limits the available bandwidth. Measuring available bandwidth is not only for knowing the network status, but also to provide information to network applications on how to control their outgoing traffic and fairly share the network bandwidth.

Another related throughput metric is the Bulk-Transfer-Capacity (BTC). The BTC of a path in a certain time period is the throughput of a bulk TCP transfer, when the transfer is only limited by the network resources and not by limitations at the end-systems. The intuitive definition of BTC is the expected long-term average data rate (bits per second) of a single ideal TCP implementation over the path in question.

In order to measure different capacity metrics, the installation of special utilities \cite{iperf} is required at both ends of path. This is uncomfortable process especially for usual Internet users who try to install modern network applications like videoconference service.

For today also there are the various systems, allowing defining an available bandwidth, but they have the disadvantages, therefore search of new solutions is claimed. Among them, such as {\em iperf, netperf, pathrate, pathload} and {\em abget}, and also a number of little-known programs {\em ncs, netest, pipechar}. We will consider the cores from the above described products.

Each of the products set forth above has disadvantages. Utilities {\em Iperf, netperf, pathrate} have one feature which is their essential disadvantage. To estimate capacity of a network it is required to instal client and server parts of the program. The utility {\em abget} demands HTTP a server on the remote server and the privilege of the superuser, and as to programs {\em ncs, netest, pipechar} so they are not adapted for operation with network screens that in modern conditions does their a little used. 

At the same time these programs use algorithms of an estimation of available bandwidth, grounded on transmission the considerable quantity of packages on a data link that reduces capacity of a network suffices and demands considerable time.

In order to construct a perfect picture of a global network (monitoring and bottlenecks troubleshooting) and develop the standards describing new appendices, the modern measuring infrastructure should be installed. In Russia different measurement projects are realized in the area of networking, for example, PingER \cite{pinger} in Institute of Theoretical and Experimental Physics (ITEP), but full access to the collected data is limited for researchers. Unfortunately, current measuring area do not reflect structure of the Russian segment of a global network.

At present time powerful measurement system like RIPE Test Box is expanded \cite{ggk}. Unfortunately, this system doesn't measure the available bandwidth, but it collects the numerical values characterized the network heals like delay, jitter, routing path, etc. This data allows us to investigate the basic interdependencies of available bandwidth from basic network parameters. Our aim is to estimate the available bandwidth from the delay value, received from one point of path.

In our work we try to present the uniform model, allowing measuring all known throughput metrics. Our method is based on testing of a network by packages of the different size. Earlier such technique called Variable Packet Size (VPS) was applied in work~\cite{dow}. The VPS technique can estimate the capacity of a hop $i$ based on the relation between the Round-Trip Time (RTT) up to hop $i$ and the probing packet size $W$. 

\section{Model}

The well-known expression for throughput metric describing a ratio between a network delay and the packet size is a version of the Little's Law \cite{klein}.
 	\begin{equation}
  B=W/D
  \label{bav}
	\end{equation}	
Here $W$ is the size of transmitted packet and $D$ is the networking packet delay. This formula is ideally for calculation of available bandwidth between two network points that are connected immediately (in other words for distantion in one hop). In general case the delay value is caused by such constant network factors as propagation delay, transmission delay, per-packet router processing time, etc \cite{pdm}.

In 1999 Downey \cite{dow} for the first time has detected linear dependence of the minimum possible round trip time on the size of transferred packets.
In 2004 precise experiments by Choi et al \cite{chm} proved that the minimum fixed delay component for a packet of size $W$ is a {\it linear} (or precisely, an {\it affine}) function of its size, 
\begin{equation}
  D^{fixed}(W)=W\sum_{i=1}^h 1/C_i + \sum_{i=1}^h \delta_i
  \label{eq2}
\end{equation}
where $C_i$ is each link of capacity of $h$ hops and $\delta_i$ is propagation delay. To validate this assumption, they check the minimum delay of packets of the same size for three path, and plot the minimum delay against the packet size. 

Let $D(W)$ represents the point-to-point delay of a packet. Here we refer to it as the minimum path transit time for the given packet size $W$, denoted by $D^{fixed}(W)=\min D(W)$. With the fixed delay component $D^{fixed}(W)$ identified, we can now substract it from the point-to-point delay of each packet to study the variable delay component $d^{var}$. The variable delay component of the packet, $d^{var}$, is given by 
\begin{equation}
	D(W)=D^{fixed}(W)+ d^{var}
	\label{dvar}
\end{equation}
 
\begin{figure}
\centering
\includegraphics[height=5cm]{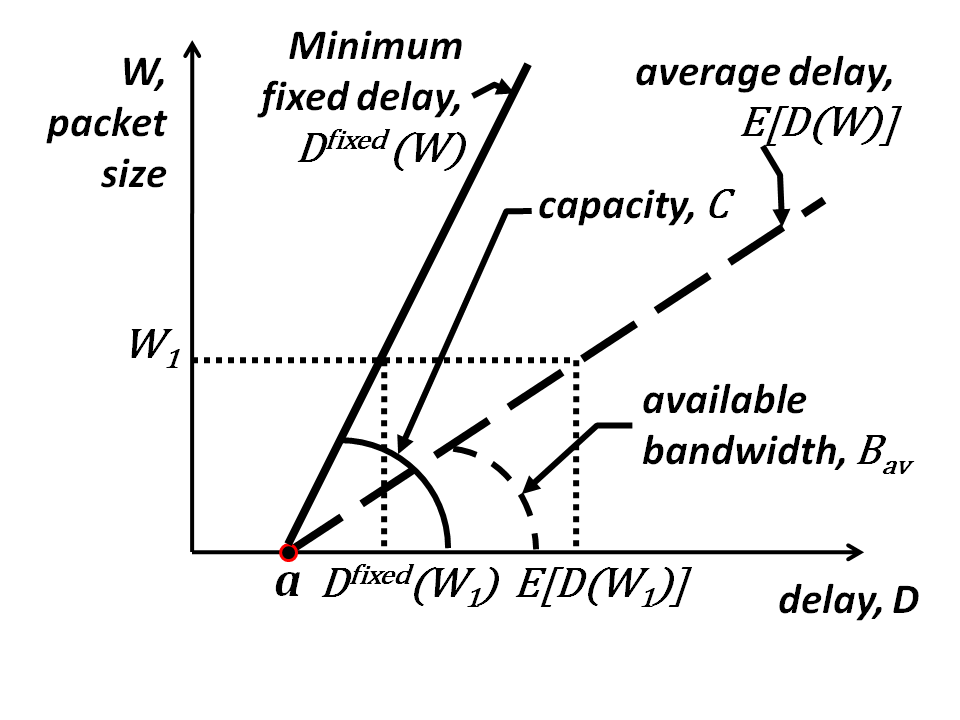}
\caption{Packet Size vs Delay}
\label{f1}
\end{figure}

On the Fig.~\ref{f1} the graphic shows the linear dependence between average network delay $D_{av}(W)=\mathbb{E}[D(W)]$ and packet size $W$ like it is constructed in paper \cite{chm}. Slope angle concerning $Y$ axe could be considered as available bandwidth $B_{av}$ in contrast to bottleneck capacity $C$ (maximum throughput) for computed minimal delay $D^{fixed}(W)$:
\begin{equation}
	D^{fixed}(W)=D_{min}+W/C,
	\label{C-for}
\end{equation}
where
\begin{equation}
	D_{min}=\lim_{W\rightarrow 0}D^{fixed}(W)
	\label{Dmin}
\end{equation}
 
Prolongation of line $D(W)$ from Fig.~\ref{f1} to $Y$ axe gives the intercept value $a=\sum_{i=1}^h \delta_i$. Then the Equation~(\ref{bav}) for the throughput metric which path consists of two or more hops should be modernized to the following view:
 	\begin{equation}
    B_{av}=W/(D_{av}-a)
    \label{Bnew}
	\end{equation}	
The value $a$ is related to the distance between the sites (i.e. propagation delay) and per-packet router processing time at each hop along the path between the sites \cite{cml,cc}. This value represents as the minimum delay $D_{min}$ for which the very small package can be transmitted on a network from one point in another. In the general case $a(n,l)$ could be considered as the linear function depended on $n$ and $l$, 
\begin{equation}
  a=f(n,l)\approx	\alpha n + \beta l
  \label{appra}
  \end{equation}
where $n$ is the number of hops (routers) that is measured by the traceroute utility and $l=\sum_n l_n$ is the sum of single length of routing path. 

The Equation~(\ref{Bnew}) gives us the simple way for estimation of throughput metrics including active bandwidth $B_{av}$ and capacity $C$. Our method supposes the variation of packet size on the same path for measurement of the throughput metrics. If the testing process between two fixed points is organized by packets with different sizes $W_1$ and $W_2$ then the delay times $D_i$ get two different values. Experiments should show the identical value for available bandwidth $B_{av}$ independently from packet size $W_i$. The system from two equations with different values of variables $D_i=\mathbb{E}[D(W_i)]$ and $W_i$ is easy solved to find $B_{av}$ and $a$:
  \begin{equation}
  B_{av}=\frac{W_2-W_1}{D_2-D_1}
  \label{Bav}
	\end{equation}	
It should be noted that similar result was first time received for {\it bandwidth-dominated\/} path in classical paper of Jacobson~\cite{jac} dedicated congestion and avoidance control. 

\begin{figure}
\centering
\includegraphics[height=5cm]{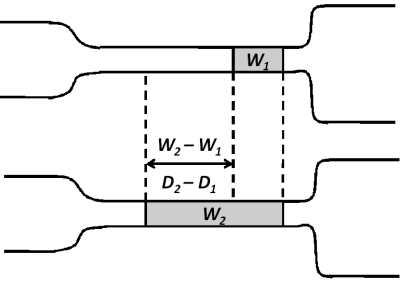}
\caption{Available Bandwidth Illustration}
\label{f2}
\end{figure}

Fig.~\ref{f2} illustrates a schematic representation of transfer of packages of the different sizes on the slowest link in the path (the bottleneck). The vertical dimension is bandwidth, the horizontal dimension is time.

Another result for capacity $C$ will turn out, if instead of the average value $D_{av}(W)$ in an analogue of the equation~(\ref{Bnew}) 
\begin{equation}
	C=\frac{W}{D^{fixed}(W)-D_{min}}
	\label{Cnew}
\end{equation}
the minimum fixed delay component $D^{fixed}(W)$ is used
\begin{equation}
	C=\frac{W_2-W_1}{D^{fixed}(W_2)- D^{fixed}(W_1)}
	\label{}
\end{equation}

It is necessary to notice that experimental definition of any throughput metrics demands carrying out of several measurements for a network delay. After these measurements are spent for packages of the different sizes, it is necessary to choose from them the minimum and average values. The minimum value will be used for calculation of available bandwidth $B_{av}$, and average value for capacity $C$. Even in work of Downey \cite{dow} it was noticed that are many data points near the minimum and we can find the minimum delay $D_{min}$ with a small number of probes at each packet size. It should be noted that the method presented in given work allows measuring the available bandwidth and capacity of the outgoing channel.

The minimal delay of datagram transmission $D_{min}$ may be calculated as
\begin{equation}
  D_{min}=\frac{W_2D_1 - W_1D_2}{W_2-W_1}
\end{equation}	
This value as well as the methods of its measurement has a important significance in applied tasks of control theory~\cite{zbp}. The second significant question of networking control theory is the distribution type for variable delay component $d^{var}$ which should be studied. To know the expression for this parameter we may easy calculate the duration of buffer for streaming aplication on receiving side.

\section{Precise Experiments}

A number of measurements in a global network have been spent for acknowledgement of our method. In this work the very first results which are already processed are presented only.
		
For practical realization of our method the sizes  $W_1$ and $W_2$  should different in several times, it is reasonable to choose 64 and 1064 bytes for Linux based systems, 32 and 1032 bytes for Windows correspondingly. The basic problem of experimental testing is the precise of delay measurements that is necessary for accurate result. The exact metering demands micro second precision for delay measurements; we are reaching such accuracy with help of RIPE Test Box mechanism \cite{ripe}. In order to prepare the experiments three Test Boxes have been installed in Moscow, Samara and Rostov on Don during 2006-2008 years in framework of RFBR grant 06-07-89074. Each RIPE Test Box represents a server under management of an FreeBSD operating system with the GPS receiver connected to it.

Characteristic times of investigated processes (a packet delay, jitter) have the order from 10 $ms$ to 1 $sec$, therefore is quite enough accuracy of system hours of a RIPE Test Box for their reliable measurement. At the first stage experiment between tt01.ripe.net (RIPE NCC at AMS-IX, Amsterdam) and tt143.ripe.net (Samara, SSAU) have been made which included
\begin{itemize}
	\item 
	Precision measurement of packet delay in the size 100 and 1100 bytes with accuracy 2-12 $\mu s$
	\item
	Measurement of available bandwidth by means of utility {\em iperf}~\cite{iperf}
	\item
	Measurement of bandwidth by a method of downloading of a file on FTP
\end{itemize}
Thus, at us it will be generated alternatively measured three sizes of throughput metrics for the subsequent comparative analysis.

It is necessary to notice that the utility {\em iperf} is started with an option {\em -u} and measures speed of a stream between two points that precisely enough corresponds to available bandwidth. Speed of downloading on ftp measures a Bulk-Transfer-Capacity (BTC) and gives strongly underestimated value. Unfortunately, at the given stage we could not spend more exact measurements, but further we assume to find partners for installation of exact utilities.

The design of the RIPE TTM system meets all requirements shown by our method, namely it allows to change the size of a testing package and to find network delay with a split-hair accuracy. 

By default, testing is conducted by packages in the size of 100 byte, but there is a page corresponding to point of the menu «Configuration» of local Test Box. On which it is possible to add testing packages to RIPE Box up to 1500 byte in size with demanded frequency. 

In our case it is reasonable to add testing 1100 (1024) byte packages with frequency of 60 times in a minute. It is necessary to notice that the results of tests will be available on next day.

Testing results are available in telnet to RIPE Test Box on port 9142. It is important to come and write down simultaneously the data on both ends of the investigated channel, in the case presented here it is tt01.ripe.net and tt143.ripe.net. Obtained data will contain required delay of packages of the different sizes. Also, we need to distinguish packages.

Therefore at first it is reversible to sending Box and we will find lines, see Table~\ref{sb}.

\begin{table*}
	\centering
		\begin{tabular}{lllllllllll}
SNDP & 9 & 1240234684 & -h & tt01.ripe.net  & -p & 6000 & -n & 1024 & -s & 1039148464\\
SNDP & 9 & 1240234685 & -h & tt164.ripe.net & -p & 6000 & -n & 100  & -s & 1039148548\\
SNDP & 9 & 1240234685 & -h & tt01.ripe.net  & -p & 6000 & -n & 100  & -s & 1039148557\\
		\end{tabular}
	\caption{The data of sending box}
	\label{sb}
\end{table*}

Last value in string is sequence number of the packet. It is necessary to us to find this number on the receiving side at the channel. The string example on the receiving side is lower resulted, see Table~\ref{rb}.

\begin{table*}
	\centering
		\begin{tabular}{lll}
		{\small
RCDP  12  2  89.186.245.200  60322  193.0.0.228  6000} & {\small 1240234684.785799} & {\small 0.044084  0X2107  0X2107  1039148464    0.000002  0.000008}\\
{\small
RCDP  12  2  89.186.245.200  53571  193.0.0.228  6000} & {\small 1240234685.788367} & {\small 0.043591  0X2107  0X2107  1039148557   0.000002   0.000008}\\
		\end{tabular}
	\caption{The data of receivig box}
	\label{rb}
\end{table*}

For set number of a package it is easy to find network delay, in our case it makes 0.044084 seconds. The following package 1039148557 has the size of 100 bytes and its delay makes 0.043591 seconds. Thus, the difference will make 0.000493 second. 

Our model assumes operations with minimal and average values; therefore we should note average values, not less than five pairs for the delay, going consistently. In our case, average difference $\mathbb{E}[D(1024)-\mathbb{E}[D(100)$ is 0.000571 seconds. (tt143 -> tt01). Then the required bandwidth of the link (tt143 -> tt01) can be calculated as

\begin{equation}
  B_{av}(tt143 \rightarrow tt01)=\frac{924\times 8}{0.000571} = 12.9 [Mbps]
  \label{t7}
\end{equation}

The minimal and average values of the return link (tt01 -> tt143) are $\mathbb{E}[D(1024)]-\mathbb{E}[D(100)]=0.000511$ second and $ D^{fixed}(1024)- D^{fixed}(100)= 0.000492$ second/ Then available bandwidth and capacity can be calculated as
\begin{align}
	C (tt01 \rightarrow tt143)=\frac{924\times 8}{0.000492} = 15.0 [Mbps]
   \label{t8}\\
  B_{av} (tt01 \rightarrow tt143)=\frac{924\times 8}{0.000511} = 14.7 [Mbps]
   \label{t9}
\end{align}

The main problem of the offered method consists in understanding, what value is measured. Actually, it can be bulk transport capacity or available bandwidth. Alternative measurements of the given values are necessary for specification.

It is ideal to compare the width received by our method to the values measured by alternative methods, first of all by means of the utility iperf. Unfortunately, such tests are not spent yet, we allocate only in the speed of FTP downloading. It makes 3.04 - 3.20 Mbps in a direction from tt143.ripe.net to tt01.ripe.net and 3.2-3.3 Mbps in the opposite direction. That is additional researches for which carrying out partners are required are necessary.

It should be noted that Table II from paper \cite{chm} gives us these values; calculated slope is inverse value to end-to-end capacity. The corresponding capacities for data set 1, 2, 3 (path 1 and 2) are {\em 285 Mbps, 128 Mbps, 222 Mbps} and {\em 205 Mbps}.

\section{AvBand Utility}

Routinely the special utilities could be used for delay measurements; we tried to test traditional {\em ping}, the new {\em UDPping} and other utility. In result of test the simplest utility {\em ping} was found to be a best choice for delay measurements. 

Utility {\em AvBand} (Available Bandwidth) has been developed, realizing the above described method, using in the basis algorithm ping. This algorithm has been developed by Mike Muus in 1983 in the USA for operating system BSD~\cite{ping}. Its advantage consists that it is possible to work with any router or the host which responds to packages of inquiries ICMP Echo. The given version of the utility is developed for platform Windows and uses library ICMP Windows API. In the near future we plan working out of the utility for Unix systems, first of all for family Linux.

\begin{figure}
\centering
\includegraphics[height=5cm]{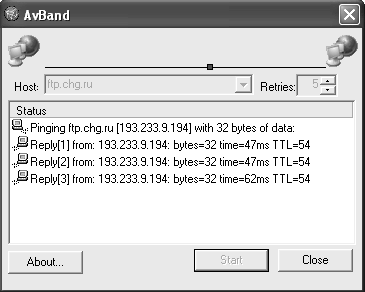}
\caption{The AvBand Screenshot}
\label{f3}
\end{figure}

The given utility defines available bandwidth of outgoing channel between host from which measurement and a remote server interesting us is spent. For this purpose the program measures RTT (Round Trip Time) that is the time between sending of inquiry and answer reception. Thus at first packages in 32 bytes (standard Windows size) are generated and their RTT is defined, and the following step forms packages of the size in 1032 bytes and is measured their RTT. On Fig.~\ref{f3} the screenshot of the program is presented.

In the field ``Host'' it is entered a host name, available bandwidth to which we are going to measure. In the field ``Retries'' the quantity of the echo-inquiries which will be sent on a remote host is underlined. After that enough to press button``Start'' and the utility will send the set quantity of packages of the size of 32 bytes, further the same quantity of packages in the size of 1032 bytes. The collected values of the received delays on each of groups of packages are averaged, and then by means of our model the available bandwidth of the channel pays off and is displayed. It is necessary to notice that the available bandwidth of the outgoing channel is measured.

\begin{table*}
	\centering
		\begin{tabular}{|c|c|c|c|c|}
			\hline
			\multicolumn{2}{|c|}{Hosts}	&	\multicolumn{3}{|c|}{Available bandwidth}	\\ \hline
			testing & remote & {\em ping} or& {\em FTP} & {\em Iperf} \\ 
			server & host     & {\em AvBand}     &    &       \\  \hline
			SSAU   & IOC RAS  & 20-20.6 & 17.6-27.4 &  \\
				     &					& {\it Mbps} & {\it Mbps} & \\  \hline
		  SSAU   & server2.hosting.reg.ru & 1150 & 1140 & \\
		  &                    & {\it Kbps} & {\it Kbps} & \\  \hline
		  OSU    & SSAU     & 2500 &  & 2450  \\
				  &                    & {\it Kbps} & & {\it Kbps}  \\  \hline
		 AIST    & SSAU     & 536 & 600 & 659  \\
				  &                    & {\it Kbps} & {\it Kbps} & {\it Kbps}  \\  \hline
		Infolada & SSAU     & 346 & 374 &   \\
		  &                    & {\it Kbps} & {\it Kbps} & \\  \hline
		VolgaTelecom    & SSAU     & 274 &  & 283  \\
			&                    & {\it Kbps} &  & {\it Kbps}  \\  \hline

		\end{tabular}
	\caption{Experimental results}
	\label{exr}
\end{table*}

For check of utility AvBand a series of experiences with use of following measuring mechanisms also has been spent: 
\begin{itemize}
	\item 
	Utility {\em AvBand}
\item 	
	Standard {\em ping}
	\item 
	{\em Iperf}
	\item 
	{\em FTP}
\end{itemize}

Measurements with Samara State Aerospace University (SSAU), Institute of Organic Chemistry of the Russian Academy of Sciences (IOC RAS), control centre RIPE in Amsterdam (RIPE),  Ohio State University (OSU), and also a number of local experiments with use of networks of various Internet Service Providers of the Samara region (Infolada, AIST, VolgaTelecom, etc.) have been currently spent. All data on experiments is resulted in the table more low.

As a case in point ADSL connection in Samara region could be chosen for illustration of our approach. The delay measurements give $D_1=18$ $ms$, $D_2=42$ $ms$, that corresponds to {\em 350 Kbps} of available bandwidth. During FTP session the delay grows to {\em 300 ms} and {\em 425 ms} that corresponds approximately to {\em 60 Kbps} of available bandwidth. This is very rough computation, but it could be made quickly and independently.

\section{Conclusion}

Now measurements are not completed yet, is planned to type the data from not less than 50 points scattered on territory of a planet. From these measurements not less than 10 should be fulfilled with application of RIPE Test Boxes. Thus, summing up to the done operation, it is possible to draw the main output: the theoretical model of calculation of an available bandwidth proves to be true.

Further it is planned to continue researches to establish type of distribution for a network delay. At definition of type of distribution it is supposed to use analogy to molecular physics, namely about distribution of molecules in the speeds Maxswell. Probably, in our case required distribution should be presented in the form of product of normally (Gaussian) distribution and the inverse function defined by the Equation~\ref{Bnew}. The knowledge of density of distribution in TCP/IP networks will help to find a new class the decision in the networked control systems.

In summary we would like to express special gratitude of Prasad Calyam and Gregg Trueb from Ohio State University for the invaluable help at carrying out of measurements. Also it would be desirable to thank all collective of technical service RIPE ncc and especially Ruben van Staveren and Roman Kalyakin for constant assistance in comprehension of subtleties of a measuring infrastructure.

\end{document}